
\pubnum{92-47}
\doublespace

\titlepage
\voffset=24pt
\title{\bf COLLECTIVE DIPOLE OSCILLATIONS IN ATOMIC NUCLEI AND SMALL METAL
PARTICLES}
\author{R. S. BHALERAO$^{1,*}$ and MUSTANSIR BARMA$^{1,2,*}$}
\address{\it
$^1$ Theoretical Physics Group \break
Tata Institute of Fundamental Research \break
Homi Bhabha Road, Colaba, Bombay 400 005, India\break
and \break
$^2$ Department of Theoretical Physics \break
University of Oxford \break
1, Keble Road, Oxford OX1 3NP, U. K.}

\vfill

\noindent PACS Nos 24.30.Cz, 25.20.-x, 36.40.+d, 78.40.Kc.


\noindent $^{*}$ e-mail addresses:
Bhalerao@tifrvax.bitnet, Barma@tifrvax.bitnet
\endpage

\abstract{The systematics of photon absorption cross sections
in  nuclei and small metal particles  are  examined as a
function of  the  number  of constituent fermions $A$.  It   is pointed
out  that  the shell-structure-linked oscillations  in  the full width at
half  maximum (FWHM) of the photoneutron cross section in nuclei,
earlier recognized for $A > 63$, in fact persist down  to the lightest
nuclei. Averaging over the oscillations or focusing on
the lower envelope of the oscillating curve (magic nuclei), the FWHM is
seen to generally decrease with increasing $A$, consistent with
$A^{-1/3}$, a dependence which was earlier known to hold
in metal particle systems. If the FWHMs are scaled by the respective
Fermi energies and the inverse radii by the Fermi wave vectors, the two
data sets become comparable in magnitude.  A schematic theoretical
description of the systematics is presented.}
\endpage

\noindent {\bf 1. Introduction}

\medskip
             It is becoming increasingly clear that there are certain
points of strong resemblance  between  the electronic properties  of
metal particles or clusters,   and  the  properties  of  atomic
nuclei (Sugano 1987). Electrons in metal particles  and nucleons in nuclei both
constitute  finite Fermi systems with temperatures much less than the
respective Fermi energies -- a fact which cuts across the very different
scales of length, mass  and energy in the two systems.  Metal particles,
like nuclei (and unlike atoms),  exhibit  saturation,  or  constancy  of
particle density, with increasing size.  Also, shell structure in energy
levels -- long familiar in  nuclear  physics -- manifests itself in  the
relative abundances, polarizabilities and ionization potentials of metal
particles as well (de Heer {\it et al} 1987).

             Here  we focus on another aspect of the  analogy,  namely,
the response of metal  particles and nuclei to electromagnetic  radiation.
In both systems, the  field resonantly excites a collective  dipolar
mode, and the wavelength of the radiation at
resonance far exceeds the size of the system.
In nuclei, it is the  giant  dipole resonance (GDR), in which
protons  are displaced with respect to neutrons and strong interactions
provide the  restoring force
(see, {\it e.g.},
Berman and Fultz 1975, van der Woude 1987). In metal particles,  the  Mie
resonance  involves  displacing  the conduction electron cloud with
respect to the background of positive ions, and there are electromagnetic
restoring forces (see, {\it e.g.},
Born and Wolf 1970, Apell {\it et al} 1990, Kresin 1992).
In this paper,  we explore
the systematics of these two  sets  of data with varying  numbers of
fermions.

     In Section 2, we discuss the available data for both nuclei and
metal particles, and draw
some empirical conclusions regarding systematic
trends, with an emphasis on the FWHMs.
This is our main result. In Section 3, we
present a schematic  theoretical description for the size dependence of
the width. A brief account of this work has already appeared (Barma and
Bhalerao 1992).
\endpage
\noindent {\bf 2. Cross-section systematics}
\medskip
\noindent  2.1 Nuclei

             Photo-neutron cross sections have been measured in a large
number of nuclei, and plots of the measured cross sections as a function
of incident photon energy, $\sigma (E)$, have been compiled
by Dietrich and Berman 1988.
For spherical nuclei,  the  cross  section for the photoexcitation of the
GDR  has a single maximum, while for deformed nuclei there is more than
one maximum (see, {\it e.g.}, Berman and Fultz 1975).
In the former case, the peak frequency  $\omega_0$ is
known to exhibit a systematic
empirical dependence on the mass number $A$: $$
\omega_0 = 31.2 A^{-1/3} + 20.6 A^{-1/6}~{\rm MeV}.  \eqno (1) $$ We shall
discuss this dependence ({\it vis-\`a-vis} the $A$ independence of
$\omega_0$ in metal particles) in Section 3.1.
The FWHM provides a simple, single characterization of the resonance
spectrum, and has been used earlier to extract global trends with varying
$A$, for heavier nuclei.  For instance, Berg\'ere (1977) and Snover (1986)
have shown plots of the FWHM in the regions $A > 90$ and $166 > A > 63$,
respectively.  These plots show that the FWHM exhibits systematic
oscillations in the ranges studied, with local minima near spherical,
near-magic nuclei.

Instead of FWHM, an alternative characterization of the data is to fit
one or two Lorentzians  to $\sigma (E)$, and thus extract the  values of
the resonance energies $\omega_{01}$ and $\omega_{02}$, the
widths $\Gamma_1$   and  $\Gamma_2$,  and   cross
sections $\sigma_1$ and $\sigma_2$,  for  the lower and higher  energy
resonances respectively.  This procedure  is usually applied to $A~>~50$,
and the  values of $\Gamma_1$ and $\Gamma_2$ are given by Dietrich and
Berman (1988). We
have observed that if the width corresponding to the larger cross section
is plotted versus $A$, then  the  result shows trends similar to those
exhibited by the FWHM. However, the FWHM has the added advantage that it
can be used across the periodic table.

              We wanted to see whether the systematics observed earlier
for the FWHM of heavier nuclei (Berg\`ere 1977, Snover 1986) persisted
in lighter nuclei as well.  We examined the FWHM in about 120 nuclei
ranging from $^3 He$ to $^{239} Pu$, using primarily the cross-section
data compiled by Dietrich and Berman (1988).
(We re-examined the heavier nuclei
in order to have a uniform procedure for all $A$.) For  those  cases where
the data follows a  curve  with  a single peak, it was straightforward to
determine the FWHM.  For cases with two or more (closely
overlapping) peaks, we
found the FWHM by drawing a smooth curve with a  single maximum through
the data points, trying to ensure that the  areas under the smooth curve and
the experimental data were nearly equal.  Those  nuclei where the data seem
incomplete ($^3H,~^{19}F$) or have too much structure ($^{14}C,
{}~^{18}O,~^{24,26}Mg$)  were  ignored.  Results  are displayed in figure
1. For light nuclei (see inset in figure 1), we also estimated the errors in
the FWHMs, arising from (a) the existence of more than one data set in
some cases, and (b) the inherent uncertainty in extracting the FWHM by our
procedure.   In the
range $A > 90$, our values agree well with those of Berg\'ere (1977). A
determination of the FWHM  using least-square fits to  a smooth
single-peaked curve would be  more  rigorous, but we do not expect it to
change our estimated values enough to affect our conclusions.

Examination of the results in figure 1 for light nuclei ($A<
50$)
(see the inset) shows that the FWHM continues to display local minima at,
by and large, the magic numbers.
The rapid
oscillations of the FWHM versus $A$ are due to the relative
crowding in of magic numbers for small $A$.
With the sole
exception of  $^{28}Si$,  all  the  minima occur at or  near  the  magic
numbers.$^{\# 1}$ Conversely, each  magic  number  has  a   corresponding
minimum, with the possible exception of $N = 40$ ($A = 72$), where there
is a hint of a local minimum,  but  the data does not allow us to draw  a
firm conclusion.  In any case,  40  is known to be a weak magic number.

The systematic oscillations in the region $A<50$  in figure 1 are
statistically significant.
As is evident from the inset, the error in the FWHM is
less than 1 MeV in almost all cases, and is generally much smaller.
The amplitude of oscillations, on the other hand, is at least 5-6 MeV
({\it e.g.}, $^3He$ to $^4He$, or $^{40}Ca$ to $^{45}Sc$), and is sometimes as
large as 14 MeV ({\it e.g.}, $^9Be$ to $^{14}N$).
The oscillations are  as systematic and as pronounced as those for
large $A$, the only difference being that there are fewer points per
oscillation.

That the photo-response of a nucleus even as light as $He$ can be thought
of in the same terms as that of heavier nuclei may seem surprising, but
the very fact that the widths for light nuclei fit in well with the
systematics across the periodic table provides an {\it a posteriori}
justification for the use of the FWHM even for $A<50$.

             An interesting  feature  of figure 1  is  the  overall downward
trend of the oscillatory  curve, evident if, for instance, we focus on
points in the lower envelope of the curve. These points correspond mostly
to spherical, magic nuclei. The manner in which the width decreases as a
function of size is discussed  below, in connection with
figure 2.
\medskip
\noindent  2.2  Metal particles

     In metal particles, the Mie resonance corresponds to the excitation
of a surface plasmon.
For a spherical particle, within the free electron
approximation, the resonance frequency $\omega_0$ is $\omega_p/\sqrt 3$, where
 $\omega_p$ is
the plasma frequency (Born and Wolf 1970).
Since $\omega_p$ depends primarily only on the
electron density, $\omega_0$ does not vary very strongly with size \dash
less than $10\%$ as the size is decreased from $\sim 100 \AA$ to $\sim
10 \AA $ (Kreibig and Genzel 1985).

             For optical absorption experiments on small  metal particles,
it is important to distinguish between two types of samples: (i) free
metal clusters  in  which size separation is  achieved  by  mass
spectroscopy, and
(ii)  metal clusters embedded in  various  matrices,
such  as  glass  or  solid  argon.  Particles are isolated  from each
other, but some spread in size  cannot be avoided.  A typical spread in
radius is $\sim 20\%$.

             Experiments on samples of type (i) have been performed  on
small metal clusters     with    between   2   and    40    conduction
electrons (Selby {\it et al} 1989, 1991, Tiggesb\"aumker {\it et al} 1992).
The data indicates that, as with nuclei, there is a
strong response  over a relatively narrow frequency  interval  in the
case  of  magic  numbers, and  over  a  much  broader frequency  range in
cases which fall between magic numbers.  In the latter case, the line
shows splittings, which can be interpreted in terms of shape
deformations.

             A larger range of sizes ($\sim 10 \AA$ to $\sim 100 \AA$)  has
been investigated (Kreibig and Genzel 1985)
in experiments on samples of type  (ii).  No
noticeable oscillations  in  $\Gamma$  versus radius  $R$  have  been
observed with such samples, but this is probably because (a) unlike for
nuclei, oscillations are averaged out due to  the distribution of sizes,
and (b) the amplitude of shell-structure-linked oscillations is expected
to decrease with increasing size, and thus be small for the above range of
sizes.  However, these experiments do reveal a  systematic dependence of
the averaged FWHM $\Gamma_{av}$ on $R$: $$ \Gamma_{av} = K ~{\hbar v_F
\over R} + \Gamma_\infty, \eqno (2) $$ where  $v_F$ is the Fermi velocity,
$K$ is a  constant of order unity and $\Gamma_\infty$ is the width in the
bulk medium.  Equation (2) describes the variation of the linewidth of
$Ag$ particles in a variety of host  matrices.  The constant $K$ depends
on  the matrix (Kreibig and Genzel 1985);
it goes down  by a factor $\sim 3$ as the matrix is
changed  from glass to an inert element  solid like $Ar$ and $Ne$,
presumably due to surface effects.

\endpage
\noindent 2.3  Similarities and differences between nuclei and metal
particles

We  wanted  to  see if (2),  which  holds  for  metal particles
with a spread in sizes, also
describes the  downward trend of $\Gamma$ in nuclei with  increasing  $A$,
evident in figure 1.
A similar $1/R$ dependence has been discussed earlier
(Myers {\it et al} 1977) for nuclei in the
range $A > 50$.  On dividing  across by the Fermi energy
$\epsilon_F$, we see that (2) predicts that $\Gamma_{av}/\epsilon_F$
is a linear function of $(k_F R)^{-1}$, where $k_F$ is the Fermi wave
vector.  Interestingly,
on  using these dimensionless  scaled variables, we can  directly
compare the $Ag$-particle and nuclear data (figure 2)
which in absolute terms differ by six orders of magnitude in
energy and five orders of magnitude in length.
We used the values
$\epsilon_F = 38$ MeV and $k_F = 1.36$ fm$^{-1}$ for nuclei, and
$\epsilon_F = 5.49$ eV and $k_F = 1.20 A^{-1}$  for $Ag$ particles.  We
have chosen to plot  data for $Ag$ particles in argon and neon matrices as
interactions  with surrounding inert gas atoms are  likely to  be
minimal, and a large range  of sizes has been studied for $Ag/Ar$
(Charl\'e {\it et al} 1989).
We have used RMS radii $\tilde R$,  as  these are well determined for
nuclei;  for  $Ag$ particles, we took $\tilde  R$ to be given by $\sqrt
{3/5}$ times the  quoted radii.  Since $\Gamma$  oscillates as a function
of size in nuclei,  and we are interested in  displaying  the overall
downward trend, we have replotted points corresponding to singly or doubly
magic nuclei from the lower envelope  of  the  curve in figure 1; the line
marked `magic' is the best fit line through these points. Thus these
points are  consistent with  a linear dependence on $(k_F \tilde R)^{-1}$,
though other monotonic variations with $\tilde R$ cannot be ruled out.  We
also examined the average downward trend of the oscillatory curve in figure
1, and found that it could also be fit to a linear dependence.  The slope
of the average line (marked `average' in figure 2) is larger than that of
the solid line, and is comparable to the slopes of the dashed and dotted
lines.  Thus (2) holds to a good approximation for nuclei also.  In
particular, it is interesting to see how well the doubly magic nuclei
$^4He,~ ^{16}O,~ ^{40}Ca,~ ^{90}Zr,$ and $^{208}Pb$ follow a straight
line.


Of course,
there are also some differences between nuclei and metal particles. We
have already pointed out a difference as regards the $A$ dependence of
$\omega_0$. The magic numbers in the two cases are also not  the
same for large $A$, because of the difference in the strengths of the
spin-orbit force in the two systems. Finally, the intercepts on the
$\Gamma/\epsilon_F$ axis, for the two sets of data in figure 2, are quite
different. The significance of this will be discussed in the following
section.
\bigskip
\noindent {\bf 3. Discussion}
\medskip
     In this section, we will give a schematic theoretical description of
the expected systematics of the resonance with size, and see how it
accords with the trends seen in experiments.

     Let us begin by recalling the principal conclusions of the
comparison between nuclei and metal particles.
First, the
resonance peak frequency $\omega_0$ exhibits a systematic variation with
$A$  for nuclei (see (1)), whereas it is roughly size independent for
metal particles. Second, the FWHM exhibits a general downward trend with
increasing $A$, consistent with an $A^{-1/3}$ dependence (figure 2). A
similar dependence is also seen in metal particle systems with a spread in
sizes but with smaller intercepts on the $\Gamma/\epsilon_F$ axis.  Third,
the FWHM exhibits strong shell effects in nuclei, and similar tendencies in
separated metal clusters. In metal particle samples with a spread in
sizes, shell effects are not seen.

\noindent 3.1  Resonance frequency

The size dependence of the natural frequency of vibration $\omega_0$ can
be deduced by using
a simple classical picture of the collective mode.
In the metal particle, the restoring force arises
from the electric field produced by layers of opposite charges on the two
sides (Ashcroft and Mermin 1976)
and acts on each of the $A$ conduction electrons in
the particle. The oscillator frequency $\omega_0$
is given by the square root of the ratio of the
total restoring force per unit displacement to the mass involved. Since
both force and mass are proportional to $A$, the frequency $\omega_0$ is
roughly size independent.
 In the nucleus, on the other hand, the restoring force arises from
short-range strong interactions amongst nucleons. In a hydrodynamic
description,  it is modelled by
the surface or volume symmetry energy terms in the
semiempirical mass formulas. In the Goldhaber-Teller model
(Goldhaber and Teller 1948), the
collective state corresponds to the motion of the proton cloud through the
neutron cloud without mutual distortion. The restoring force is
proportional to $A^{2/3}$ and the mass parameter is proportional to $A$.
Hence, $\omega_0 \sim A^{-1/6}$. In the Steinwedel-Jensen model
(Steinwedel and Jensen 1950), on
the other hand, the relative proton-neutron density changes in such a way
as to maintain constant overall density  throughout. The restoring force
per unit mass is proportional to $R^{-2}$, and hence $\omega_0 \sim
A^{-1/3}$.

\noindent  3.2  Resonance widths \dash overall trend

     Turning to the FWHM in nuclei, one may distinguish
between two types of contributions. Firstly, there is the intrinsic width
of the resonance which comes from the finite lifetime of the collective
mode, and which is present in all cases. The intrinsic width itself
receives contributions from a variety of physical mechanisms to be
discussed below.  This is the only contribution to the FWHM in spherical
nuclei.  Secondly, in nonspherical nuclei, static deformations in shape
can lead to two distinct resonance frequencies, corresponding to a
splitting of the line. In such cases, the FWHM receives additional
contributions.

     The intrinsic width $\Gamma_i$ can be written as the sum of three
terms (see, {\it e.g.}, van der Woude 1987) $$ \Gamma_i =\Delta\Gamma +
\Gamma^{\uparrow} + \Gamma^{\downarrow} , \eqno(3) $$ reflecting
contributions from distinct physical effects.
 The fragmentation width $\Delta\Gamma$ corresponds to the fact that
the collective $(1p-1h)$ state which is the
doorway state for the GDR, is not a single state, but is in most cases
already appreciably fragmented. This effect (mean-field
damping or one-body friction)
is the finite nucleus analogue
of Landau damping in a bulk medium.
 It occurs due to the scattering of the nucleons from
the `wall' or `surface' of the self-consistent mean field potential.
The second term
 $\Gamma^{\uparrow}$ is the escape or decay width corresponding to the
direct coupling of the $(1p-1h)$ doorway state to the continuum, giving
rise to its decay into a free nucleon and an $(A-1)$ nucleus.
 Finally, the spreading width $\Gamma^{\downarrow}$ is due to the coupling
of the $(1p-1h)$ doorway state to more complicated $(2p-2h)$ states of the
nucleus, the transition occurring on account of genuine two-body effects
(collisional damping or two-body friction).

Let us see how each contribution to (3) is expected to vary
with radius $R$. Our arguments are schematic, and aimed at establishing
the general, systematic trend with size.

      The $R$-dependence of the first two terms may be
estimated
using a simple argument based on estimating the frequency of
collisions with the surface. Such an argument has been used successfully
(Kreibig 1974) to estimate $\Delta\Gamma$ in metal particles; the
estimate agrees with the result of a quantum calculation
(Kawabata and Kubo 1966, Barma and Subrahmanyam 1989, Yannouleas and
Broglia 1992) of the
fragmentation width, within a square-well model.
The idea is that individual fermions moving with Fermi velocity
$v_F$ hit the
wall with mean time $\sim R/v_F$; the inverse time $\sim v_F/R$ then
determines the contribution to the width arising from wall effects ---
both for $\Delta\Gamma$ and $\Gamma^{\uparrow}$.
The subject of one-body dissipation has also been
discussed extensively in the nuclear physics literature
(Brink 1957, Blocki {\it et al} 1978, Yannouleas 1985).
%
%
The width $\Gamma^{\downarrow}$,
on the other hand, arises from two-particle collisions.
We expect $\Gamma^{\downarrow}$ to vary smoothly with energy and nuclear
size for magic cases, since the collisional mean free path $\lambda$ shows
similar smooth variations (see, {\it e.g.}, Wambach 1988).
The average time between two collisions
is $\sim\lambda/v_F$ and hence $\Gamma^{\downarrow}$ is expected to be
$\sim v_F/\lambda$. This is the only contribution in the
$R\rightarrow\infty$ limit.

The total intrinsic width is thus expected to be $$ \Gamma_i
=\Gamma_\infty^{\downarrow} + c \hbar v_F/R,  \eqno(4) $$
where $\Gamma_\infty^{\downarrow}$ is the spreading width in the
bulk limit, and $c$ is a constant. This is in
agreement with (2) and the data presented in figure 2 for spherical,
magic nuclei.

It is interesting to consider the limit $R\rightarrow\infty$,
corresponding to nuclear matter or the bulk metal.  From figure 2, we see
that if the straight lines are extrapolated towards $(k_F{\tilde
R})^{-1}=0$, the resulting intercept on the $\Gamma/\epsilon_F$ axis is
much larger for nuclei than for metal particles. This is not too
surprising, as $k_F \lambda$ is much smaller in nuclear matter
(Wambach 1988) than
in bulk metals at room temperature
(Ashcroft and Mermin 1976), reflecting the greater effect
of collisions in the former case.  The limiting contribution
$\Gamma_\infty^{\downarrow}$ constitutes a substantial fraction of the
total width for spherical nuclei; for $^{208}Pb$ it is about $50\%$.
$^{\# 2}$
On the other hand, for metal particles of comparable $A$, the
contribution
  $\Gamma_\infty^{\downarrow}$
     constitutes a much smaller fraction of
  $\Gamma_i$.

\noindent  3.3  Resonance widths \dash shell effects

     So far we have discussed only the monotonic variation of the FWHM
that obtains for spherical nuclei. As we see from figure 1, when we consider
$all$ nuclei, superimposed on this monotonic variation, there are striking
and strong shell-structure linked oscillations in the FWHM.  (As mentioned
in Section 2.2, the FWHM in metal clusters also seems to show similar
tendencies.) We discuss the origin of these oscillations in two broad
representative regions, namely $150<A<190$ and $80<A<150$.

     In the range $150<A<190$, Dietrich and Berman (1988) have fitted
two-component Lorentz curves to the photoneutron cross section data. This
indicates a splitting of the line, due to deformation of the nucleus.
As in Section 2.1, we denote the two resonance energies by $\omega_{01}$
and $\omega_{02}$, with $\omega_{01}<\omega_{02}$,
and the corresponding widths by $\Gamma_1$ and
$\Gamma_2$. For a spheroidal deformation, (1) leads us to expect that
$\omega_{01}$ and $\omega_{02}$ correspond to oscillations along the
semimajor and semiminor axes respectively. Equation (4) then implies that
$\Gamma_1 < \Gamma_2$.
This is indeed found to be true, for
$150<A<190$, for the values of the widths
tabulated by Dietrich and Berman (1988).
In fact, with the exceptions of $^{55}Mn$ and
$^{63}Cu$, this is true for all the nuclei listed there.
This provides additional evidence for the overall decrease of the
intrinsic width with increasing radius. The corresponding
cross sections  $\sigma_1$ and $\sigma_2$ also generally satisfy
$\sigma_1<\sigma_2$. The increase of the FWHM away from the spherical
cases can be ascribed, at least partially, to the fact that deformations
produce a splitting of the line, and also cause $\Gamma_2>\Gamma( >
\Gamma_1)$, where $\Gamma$ would be the width if the nucleus were
undeformed. Since deformations of the shape follow shell-structure
systematics, with smallest deformations close to the magic numbers, so
does the FWHM (Okamoto 1958).

In the region $80<A<150$, Dietrich and Berman (1988) have fitted
one-component Lorentz curves to the photoneutron cross section data (with
the exceptions of $^{127}I$ and $^{148}Nd$).
 As is clear from figure 1, oscillations of the FWHM versus $A$ in
this region are as prominent as those for $150<A<190$. Thus, even when the
line is unsplit, the width of the best-fit Lorentzian oscillates as a
function of $A$, with minima at the magic numbers. This indicates
that the intrinsic width $\Gamma_i$ can itself show shell-structure linked
oscillations.$^{\# 3}$ Berg\`ere (1977) has correlated the FWHM in
this region with the ratio $E(4^+)/E(2^+)$, which characterizes the
`softness' of nuclei. Here $E(J^+)$ is the energy of the first $J^+$ state
in the nuclear spectrum.

The theoretical considerations we have presented in this section are
schematic, and aimed at
understanding the empirical systematics observed --- in contrast
to more detailed
theoretical studies of the width for $individual$ nuclei.
\bigskip
\noindent {\bf 4. Conclusion}
\medskip
We  conclude  by recapitulating the main  points  of  this paper.  The
FWHM of the total photoneutron cross  section shows shell-structure-linked
oscillations as a function of $A$ even for $3<A<50$.  Disregarding
oscillations, for instance by focusing on  magic nuclei,  the  FWHM
generally decreases  with  increasing  $A$ approximately as $A^{-1/3}$
(see (4)).
Striking  similarities are seen when the FWHMs for nuclei are compared
with photoabsorption FWHMs in small  metal  particles,  after proper
rescaling  of  the energies and lengths.

The systematics which have been pointed out and discussed in this paper
bring the task of a theory into better focus. While a complete theory has
not been presented here, we have given a schematic theoretical description
which allows one to understand at least the principal trends. However, it
is the empirical observations which constitute the main result of this
paper.
\bigskip
\noindent  {\bf Acknowledgments}
\medskip
             We  are  grateful  to C. V. K. Baba,  R.K.  Bhaduri, V.
Subrahmanyam and C.S. Warke for very useful discussions and comments on the
manuscript, and  to B.L. Berman for bringing the work of Dietrich and
Berman (1988) to our
attention. A part of the work was done when one of us (RSB) was visiting
McMaster University, Hamilton, Canada. MB acknowledges support by the
Science and Engineering Research Council (UK) under grant no. GR/G02741.

\endpage

\noindent  {\bf References}

\noindent  Apell S P, Giraldo J and Lundqvist S 1990 {\it Phase
Transitions} {\bf 24-26} 577

\noindent Ashcroft N W and Mermin N D 1976 {\it Solid State Physics}
(New York: Holt, Rinehart and Winston) ch. 1

\noindent Barma M and Bhalerao R S 1992 in {\it Physics and Chemistry
of Finite Systems: From Clusters to Crystals} (eds) P Jena, S N
Khanna and B K Rao (Dordrecht: Kluwer) NATO ASI Sr. {\bf Vol II} p. 881

\noindent Barma M and Subrahmanyam V 1989 {\it J. Phys. Cond. Matter}
{\bf 1} 7681

\noindent Berg\`ere R 1977 in {\it Photonuclear Reactions} (eds) S Costa
and C Schaerf, Lecture Notes in Physics (Berlin: Springer) {\bf Vol 61},
ch. III, fig. 23a, p. 114

\noindent Berman B L and Fultz S C 1975 {\it Rev. Mod. Phys.} {\bf 47}
713

\noindent Blocki J, Boneh Y, Nix J R, Randrup J, Robel M,
Sierk A J and Swiatecki W J 1978 {\it Ann. Phys. (NY)} {\bf 113} 330

\noindent Born M and Wolf E 1970 {\it Principles of Optics} (Oxford:
Pergamon) ch. 13

\noindent Brink D M 1957 {\it Nucl. Phys.} {\bf 4} 215

\noindent Charl\'e K P, Schulze W and Winter B 1989 {\it Z. Phys.}
{\bf D 12} 471

\noindent de Heer W A, Knight W D, Chou M Y and Cohen M L 1987 {\it
Solid State Physics} {\bf 40} 93

\noindent Dietrich S S and Berman B L 1988 {\it At. Data Nucl. Data
Tables} {\bf38} 199

\noindent Goeke K and Speth J 1982 {\it Ann. Rev. Nucl. Part. Sci.} {\bf
32} 65

\noindent Goldhaber M and Teller E 1948 {\it Phys. Rev.} {\bf 74} 1046

\noindent Harakeh M N 1985 contribution to {\it XVII Summer School on
Nuclear Structure by means of Nuclear Reactions} Mikolajki, Poland,
unpublished

\noindent Kawabata A and Kubo R 1966 {\it J. Phys. Soc. Jpn.} {\bf 21}
1765

\noindent Kreibig U 1974 {\it J. Phys.} {\bf F4} 999

\noindent Kreibig U and Genzel L 1985 {\it Surf. Sci.} {\bf 156} 678

\noindent Kresin V V 1992 {\it Phys. Rep.} {\bf 220} 1

\noindent Myers W D, Swiatecki W J, Kodama T, El-Jaick L J and
Hilf E R 1977 {\it Phys. Rev.} {\bf C 15} 2032

\noindent Okamoto K 1958 {\it Phys. Rev.} {\bf 110} 143

\noindent Preston M A and Bhaduri R K 1975 {\it Structure of the
Nucleus} (Reading: Addison-Wesley) figs. 3.2 and 10.12

\noindent Selby K, Vollmer M, Masui J, de Heer W A and Knight W D
1989 {\it Phys. Rev.} {\bf B40} 5417

\noindent Selby K, Kresin V, Masui J, Vollmer M, Scheidemann A and
Knight W D 1991 {\it Z. Phys.} {\bf D 19} 43

\noindent Snover K A 1986 {\it Ann. Rev. Nucl. Part. Sci.} {\bf 36} 545

\noindent Speth J and van der Woude A 1981 {\it Rep. Prog. Phys.} {\bf
44} 719

\noindent Steinwedel H and Jensen J H D 1950 {\it Z. Naturforschung}
{\bf 5a} 413

\noindent Sugano S 1987 in {\it Microclusters} (eds) S Sugano, Y
Nishina and S Ohnishi (Berlin: Springer)

\noindent Tiggesb\"aumker J, K\"oller L, Lutz H O, and
Meiwes-Broer K H 1992 {\it Chem. Phys. Lett.} {\bf 190} 42

\noindent van der Woude A 1987 {\it Prog. Part. Nucl. Phys.} {\bf 18} 217

\noindent van der Woude A 1991 in {\it Electric and Magnetic Giant
Resonances in Nuclei} (ed) J Speth (Singapore: World Scientific)

\noindent Wambach J 1988 {\it Rep. Prog. Phys.} {\bf 51} 989

\noindent Yannouleas C 1985 {\it Nucl. Phys.} {\bf A 439} 336

\noindent Yannouleas C and Broglia R 1992 {\it Ann. Phys. (N.Y.)} {\bf
217} 105

\endpage
\noindent {\bf Footnotes}

\item {\# 1} Although $^{28}Si$ is not a magic nucleus, it displays
behaviour similar to a magic nucleus in at least one other context:
the plot of nuclear electric quadrupole moment vs $Z$ or $N$
passes through a zero near $^{28}Si$, indicating a prolate to oblate
transition. Similar transitions also occur at the magic numbers
(Preston and Bhaduri 1975).

\item {\# 2} For real nuclei, the full two-body contribution
$\Gamma^{\downarrow}$ may differ significantly from
$\Gamma^{\downarrow}_{\infty}$ .

\item {\# 3} This is in contrast to the statements made in various reviews
(Speth and van der Woude 1981, Goeke and Speth 1982, Harakeh 1985, van
der Woude 1991) that the width, like $\omega_0$, varies smoothly with
$A$.

\endpage
\noindent  {\bf Figure Captions}

\item  {\rm Figure 1.}  $\Gamma$ is the full width  at  half maximum of the
total photoneutron cross section on a nucleus when plotted as a function
of the incident photon energy. Compilation of the cross-section data by
Dietrich and Berman (1988) was used to determine $\Gamma$.
  Note the systematic modulations  in the
$\Gamma$ vs $A$ curve, with minima  at the proton $(Z)$ or  neutron  $(N)$
magic numbers.  The minimum at $A=28$ corresponds to $^{28}Si$; see
footnote $\# 1$. The curve is drawn as a guide to the eye.
 Dashed  lines  indicate regions of sparse or nonexistent data. The
inset shows the region $A\le 45$ in greater detail. ($\Gamma$ for $Ne, Cl$
and $K$ is obtained from data on natural samples.) The statistical
significance of the oscillations of the curve is discussed in the text.

\item  {\rm Figure 2.} $\Gamma/\epsilon_F$ vs.  $(k_F  {\tilde R})^{-1}$
for metal particles and  nuclei.  $\Gamma$ is the same as in figure 1.
The dashed and dotted lines are best
fits for $Ag/Ar$ ($+$, Kreibig and Genzel 1985, Charl\'e  {\it et al} 1989)
and $Ag/Ne$ (Kreibig and Genzel 1985). The nuclei shown
here are singly $(\bullet)$ or doubly $(\circ)$ magic nuclei from the
lower envelope of the oscillating curve in figure 1. The solid straight line
is the best fit to this data set. The dot-dashed line indicates the
`average' trend of the oscillatory curve in figure 1.  Note  the
similarities  between the  scaled  nuclear  and particle data despite the
fact that the two data sets differ by six orders of magnitude in energy
and five orders of magnitude in length.  For $A \ge 90$, not all error
bars are shown; nuclei in the same cluster have roughly similar error
bars.

\bye